\documentclass[10pt, a4paper]{article}

\usepackage{float}
\usepackage{amsmath}
\usepackage{amssymb} 
\usepackage{amsthm} 
\usepackage{mathtools, nccmath} 
\usepackage{booktabs} 
\usepackage{array} 
\usepackage{caption}
\usepackage{subcaption} 
\usepackage{makecell} 
\usepackage{titlesec} 
\usepackage{adjustbox}
\usepackage{natbib}    
\usepackage[hyphens]{url}
\usepackage{hyperref}
\usepackage{authblk} 

\theoremstyle{remark}
\newtheorem{remark}{Remark}[subsection]

\graphicspath{{./Graphs}}

\begin{document}
\title{Credibility Clustering for Experience Rating: \\ 
	\large An Automated Approach to Risk Group Allocation in Group Life Insurance}
	
	\author[1]{Thom van Rijn\thanks{Corresponding author: thom.van.rijn@uni-hamburg.de, ORCID: \href{https://orcid.org/0009-0005-1398-3370}{0009-0005-1398-3370}}}
	\author[2]{Frank Weber}
	\author[1,2]{Michael Merz}
	
	\renewcommand\Affilfont{\small}
	
	\affil[1]{University of Hamburg Business School, University of Hamburg, Hamburg, Germany}
	\affil[2]{Department of Mathematics and Computer Science, University of Basel, Basel-Stadt, Switzerland}
	
	\date{November 4, 2026}
\maketitle
\vspace*{-2.5em}

\begin{abstract}
	In insurance, experience rating plays a crucial role in mitigating adverse selection by adjusting premiums according to historical claims experience. A key step in this process is the classification of insured collectives into homogeneous risk groups.
	Within group life insurance, this classification traditionally relies on manual decisions and arbitrary thresholds, 
	which may lead to inefficiencies and inconsistent outcomes. This paper proposes a data-driven methodology for 
	automating risk group allocation based on a combination of
	the multidimensional B\"uhlmann-Straub model and k-means clustering -- referred to as credibility clustering.
	By incorporating credibility weights, the method accounts for the varying statistical significance of experience data and ensures
	that risk classifications are both actuarially sound and practically implementable. Empirical results based on real-world
	group life insurance data from a major European insurer demonstrate that the proposed framework outperforms the two k-means benchmark variants considered in the empirical application in terms of viability, temporal stability and accounting for statistical reliability.
	
\end{abstract}

\section{Introduction}
\label{sec:int}

In insurance, the risk evaluation of sub-collectives is of primary importance in the pricing 
process. 
One essential step involves the classification of sub-collectives into risk groups. The 
claims experience of the resulting risk groups has a marked impact on the experience rating, 
and, consequently on the resulting premiums of the sub-collectives. Without a proper risk 
classification, adverse selection is likely to deteriorate the overall risk pool over time. 
\par 
Depending on the constraints of the use case, various classification methods may be applied. 
Although the underlying assumptions and resulting outcomes of these methods can differ 
substantially, they all rely on historical claims data. This common foundation enables a 
formal mathematical representation of the risk classification framework, which is described 
below.
\\ \noindent  
\\ \noindent
\textbf{Mathematical formulation}
\\ \noindent
Let $I$ denote the number of sub-collectives (e.g. risks) in the insurance portfolio and $T$ 
denote the number of years for which we have historical data from the sub-collectives. For 
every sub-collective $i \in \{1,2,\dots,I\}$ and year $t \in \{1,2,\dots,T\}$ we have 
statistical information from $M$ risk measures, which we denote as the claims experience 
\begin{equation*}
	\boldsymbol{x}_{ti} = (x_{ti}^{(1)}, x_{ti}^{(2)}, \dots, x_{ti}^{(M)}) \in \mathbb{R}^M
\end{equation*}
and its corresponding volume measures
\begin{equation*}
	\boldsymbol{w}_{ti} = (w_{ti}^{(1)}, w_{ti}^{(2)}, \dots, w_{ti}^{(M)}) \in [0, \infty)^M
\end{equation*}
(cf. Table~\ref{table:setting}).
Examples of claims experience and volume measure pairs include the claims ratio and the 
associated exposure, or the observed-to-expected ratio for the number of claims and the 
number of expected claims.
\begin{table}[H]
	\centering
	\captionsetup{width=0.85\linewidth}
	\begin{tabular*}{0.8\textwidth}{@{\extracolsep{\fill}}lcccc}
		\toprule
		& \multicolumn{4}{c}{\textbf{Sub-collectives} $i$} \\
		\cmidrule(lr){2-5}
		\textbf{Years} $t$ & 1 & 2 & $\cdots$ & $I$ \\
		\midrule
		1 & $(\boldsymbol{x}_{11}, \boldsymbol{w}_{11})$ & $(\boldsymbol{x}_{12}, \boldsymbol{w}_{12})$ & $\cdots$ & $(\boldsymbol{x}_{1I}, \boldsymbol{w}_{1I})$ \\
		2 & $(\boldsymbol{x}_{21}, \boldsymbol{w}_{21})$ & $(\boldsymbol{x}_{22}, \boldsymbol{w}_{22})$ & $\cdots$ & $(\boldsymbol{x}_{2I}, \boldsymbol{w}_{2I})$ \\
		$\vdots$ & $\vdots$ & $\vdots$ & $\ddots$ & $\vdots$ \\
		$T$ & $(\boldsymbol{x}_{T1}, \boldsymbol{w}_{T1})$ & $(\boldsymbol{x}_{T2}, \boldsymbol{w}_{T2})$ & $\cdots$ & $(\boldsymbol{x}_{TI}, \boldsymbol{w}_{TI})$ \\
		\bottomrule
	\end{tabular*}
		\caption{Overview of the available statistical information in the risk classification and experience rating process.}
	\label{table:setting}
\end{table} 
The aim of risk classification is to allocate the $I$ sub-collectives into $K$ disjoint risk
groups, where $K < I$. These risk groups should collectively contain all sub-collectives. 
Additionally, they should exhibit homogeneity within groups and heterogeneity between groups. 
Depending on the granularity of the sub-collectives, risk classification comes with 
additional challenges and constraints.
\\ \noindent 
\\ \noindent
\textbf{Data granularity in risk classification}
\\ \noindent
The granularity of the sub-collectives has a large impact on the efficacy of the risk 
classification and the resulting experience rating factors. For individual insureds or similarly sized exposure units, regression-based approaches such as generalized linear or generalized linear mixed models can directly account for observable risk characteristics and repeated observations~\cite{antonio2012statistical}. 
\par
However, when the units of interest are aggregated collectives whose exposures differ substantially, the observed experience measures exhibit strongly heterogeneous statistical precision. This motivates an explicit credibility adjustment prior to clustering.
\\ \noindent 
\\ \noindent
\textbf{Use case}
\\ \noindent
This limitation can be clearly observed in the use-case of this paper; experience rating of 
mortality benefits in group life insurance, where the claims experiences 
$\boldsymbol{x}_{ti}$ take the form of observed-to-expected ratios and the volume measures 
$\boldsymbol{w}_{ti}$ equal the denominator of the claims experience (i.e. the expectation). 
Apart from statistical reliability, the group life framework presents the following 
challenges and constraints:
\begin{enumerate}
	\item Low claim probabilities: For a working population, the mortality rates lie between 
	0.01\% and 1\%. Consequently, sub-collectives need to be large to be classified into risk 
	groups with confidence. Economic sectors, which describe the economic 
	activity of the underlying employer, therefore constitute the operational granularity at 
	which experience-rating adjustments are implemented in practice. Given the low mortality 
	rates, this aggregation level also 
	provides sufficient exposure to obtain statistically reliable experience measures. 
	However, even after aggregation, economic sectors remain heterogeneous in size 
	and therefore exhibit varying levels of statistical reliability.
	\item Lack of available explanatory variables: Unlike individual-life insurance, there is 
	generally very little information on explanatory variables related to the mortality risk 
	of the insured individuals of a policy that is allowed to be used without constraints. 
	Hence, besides age and sex, the most frequently used variables for risk classification 
	and experience rating are on a policy-level (i.e. company-level), which starkly limits 
	these processes. Furthermore, due to the low claim probabilities, the addition of 
	explanatory variables -- which leads to more partitions in groups -- has a significant 
	impact on the statistical reliability of these groups. Consequently, caution is typically 
	exercised when introducing additional explanatory variables.   
\end{enumerate}
The combination of these challenges leads to non-viable solutions when directly applying risk 
classification methods. For example, the application of known clustering methods, such as 
k-means \cite{k-means} \cite{k-means_2}, within this setting leads to many sub-collectives 
being allocated to the extreme risk groups (e.g. the best risk group when sub-collectives 
have no observed mortality cases), even though these observations are not statistically 
reliable due to their low volume measures. Consequently, the risk group factors resulting from 
experience rating based on these risk groups are extreme and may not adequately reflect the 
underlying mortality risk. Additionally, this risk 
classification is unstable in the temporal dimension, as one observed mortality case in a 
sub-collective can lead to a change of allocation from the best to the worst risk group.
\\ \noindent 
\\ \noindent
\textbf{Alternative methods}
\\ \noindent
Other methods have overcome these challenges by applying agglomerative 
hierarchical clustering methods,
whereby points which statistically differ the least are merged into one 
cluster~\cite{mack1997schadenversicherungsmathematik}, or the 
clustering is performed sequentially until enough small homogeneous clusters have been 
created~\cite{yeo2001clustering}.
Although promising, these methods are less suitable for our application, because the starting 
data granularity (i.e. economic sector) is predetermined in our application and the method is 
not minimizing a mathematical objective function, hence not guaranteeing a local minimum. 
\par
Furthermore, the objective in our use case is not individual policy segmentation. Instead, 
economic sectors constitute the operational granularity at which experience information is 
available and at which risk-group adjustments are intended to be implemented in practice. 
Consequently, within-sector heterogeneity is not eliminated but forms part 
of the aggregated sector-level experience.
\par
Alternatively, the k-means distances can be weighted. 
Consequently, a proxy for the statistical reliability is chosen as the weight.
Available options include the volume measures $\boldsymbol{w}_{ti}$ or exposure measures 
related to the sub-collectives.
However, when applying these weights, the observed-to-expected ratios in 
$\boldsymbol{x}_{ti}$,
which is vital for allocating to a risk group,
will be (partially) canceled out.
This causes unfair reward and punishment for small and large data points respectively.
\\ \noindent 
\\ \noindent
\textbf{Research aim}
\\ \noindent
The lack of suitable risk classification
techniques for experience rating in group life insurance 
motivates the development
of a more robust framework that incorporates
actuarial credibility concepts into the
clustering process that aims to overcome the following 
methodological and practical challenges:
\begin{itemize}
	\item Statistical unreliability of small groups:
	Clustering methods typically treat all data
	points equally, disregarding their statistical
	credibility. In group insurance, this can
	lead to volatile and misleading results,
	particularly for small or low-volume risk
	groups with highly variable claims histories.
	\item Non-viable risk group
	factors: Clusters based on raw
	observed-to-expected ratios may result
	in extreme risk group factors that are
	difficult to justify or use in practice.
	\item Temporal instability: As new claims data become
	available, cluster assignments may change drastically over time, 
	leading to unstable pricing and potential inconsistencies in underwriting decisions.
\end{itemize}
\par
In this paper, we
introduce credibility clustering: a framework
that combines credibility theory and
clustering to derive stable and interpretable
risk groupings. The multidimensional B\"uhlmann-Straub model is used to
weight the observed experience according
to its statistical reliability and k-means clustering is
performed on this transformed and smoothed data. 
\\ \noindent 
\\ \noindent
\textbf{Paper outline}
\\ \noindent
The paper is structured
as follows:
Section 2 presents the methodology of
credibility clustering in detail.
Section 3 provides empirical results and
a comparison to alternative clustering
approaches.
Section 4 concludes with a summary and
suggestions for further research.

\section{Methodology}
\label{sec:methodology}

This section presents the methodology underlying the proposed \textit{credibility clustering} 
framework. The approach integrates two established methods:

\begin{enumerate}
	\item \textbf{Credibility theory}, which adjusts the observed claims experience based on 
	the statistical reliability of the data,
	\item \textbf{Clustering}, which groups insured entities into risk classes based on these 
	adjusted and reduced features.
\end{enumerate}

The aim is to construct risk groups that are statistically sound, practically applicable, and 
stable over time. The methodology is modular and can be adapted to different data 
granularities, time windows, and clustering resolutions.

The core of the framework is outlined in Sections~\ref{sec:credibility} 
and~\ref{sec:clustering}, beginning with a review of credibility theory.

\subsection{Credibility Theory}
\label{sec:credibility}
Credibility theory is a mathematical framework rooted in Bayesian statistics. 
It provides a tool for forecasting the claims experience of an insurance portfolio 
without imposing restrictive assumptions on the distribution of the underlying risks. 
Instead, given the risk characteristic $\Theta$, the credibility estimator is required to 
\begin{enumerate} 
	\item Take the form of an affine-linear combination of the observed claims experience 
	$(X_1, \dots, X_T)$, \[ \widehat{\mu(\Theta)} = a_0 + \sum_{t=1}^{T} a_t X_t, \] where 
	$(a_0,\dots,a_T)\in\mathbb{R}^{T+1}$.
	\item Minimize the mean squared error with respect to the conditional 
	expected claims experience \[ \mu(\Theta)=\mathbb{E}[X_{T+1}\mid\Theta], \] i.e. \[ 
	\mathbb{E}\!\left[ 
	\bigl(\widehat{\mu(\Theta)}-\mu(\Theta)\bigr)^2 \right] \rightarrow \min. \]
\end{enumerate} 
The resulting credibility estimator reflects the principle of combining the individual claims 
experience with 
the experience of the collective. As shown in Equation~\eqref{eq:cred_estimators_mult}, this 
is achieved through the credibility weight, which determines the relative influence of both 
components and depends on their statistical reliability~\cite{buhlmann2005course}.
\par
Consequently, credibility theory is particularly well suited for applications involving 
heterogeneous risk 
structures and 
limited individual exposure, such as experience rating in group life 
insurance~\cite{norberg1989experience}. 
More broadly, the framework has had a substantial impact on actuarial practice, with 
applications ranging from 
automobile bonus systems~\cite{norberg1976credibility} to individual life 
insurance~\cite{gong2018credibility}. 
\par
In this paper, we employ the multidimensional \textit{Bühlmann--Straub model}
 \cite{buhlmann2005course, jewell1973multi} to derive 
stabilized risk indicators for each group. These indicators subsequently serve as inputs for 
a clustering procedure.

\subsubsection{The multidimensional Bühlmann--Straub model}
\label{sec:mbs}
Let \( \theta_i \) denote the unobservable risk characteristic of group \( i \in 
\{1, 2, \dots, I\} \), assumed to be a realization of a random variable \( 
\Theta_i \). The observed 
claims experience \( \boldsymbol{x}_{ti} \) and volume measure \( 
\boldsymbol{w}_{ti} \) correspond 
to the vectors presented in Section~\ref{sec:int}. In the multidimensional 
Bühlmann--Straub model, \( \boldsymbol{x}_{ti} \) is modeled as a realization of 
the random vector \( \mathbf{X}_{ti}=(X_{ti}^{(1)},\dots,X_{ti}^{(M)}) \), 
representing the claims experience of risk \(i\) in year \(t\). Additionally, the 
following weight matrices are defined:
\[\boldsymbol{w}_{ti} = 	\begin{pmatrix} w_{ti}^{(1)} & \cdots & 0 \\ \vdots & 
	\ddots & 0 \\ 0 
	& \cdots &  w_{ti}^{(M)} \end{pmatrix} \]
and
\[ \boldsymbol{w}_{\bullet i}^{-1/2} = 	\begin{pmatrix} (\sum_{t=1}^{T} w_{ti}^{(1)})^{-1/2} & 
	\cdots & 0 \\ \vdots & \ddots & 0 \\ 0 & \cdots &  (\sum_{t=1}^{T} w_{ti}^{(M)})^{-1/2} 
	\end{pmatrix}. \]

The multidimensional Bühlmann--Straub model is based on the following assumptions:
\begin{enumerate}
	\item Conditionally on \( \Theta_i \), \( i = 1, \dots, I \), the random variables \( 
	\mathbf{X}_{ti} \) for \( t = 1, \dots, T \) are independent and satisfy:
	\[
	\mathbb{E}[\mathbf{X}_{ti} \mid \Theta_i] = \boldsymbol{\mu}(\Theta_i), \qquad 
	\mathrm{Cov}(\mathbf{X}_{ti}, \mathbf{X}_{ti}^{'}  \mid \Theta_i) = 
	\begin{pmatrix} \dfrac{\sigma_1^2(\Theta_i)}{w_{ti}^{(1)}} & \cdots & 0 \\ \vdots & 
		\ddots & 0 \\ 0 & \cdots & \dfrac{\sigma_M^2(\Theta_i)}{w_{ti}^{(M)}} \end{pmatrix}
	\]
	
	\item The pairs \( (\Theta_1, \mathbf{X}_1),\ldots,(\Theta_I, \mathbf{X}_I) \), where 
	$\mathbf{X}_i = (\mathbf{X}_{1i}, \dots, \mathbf{X}_{Ti})$, are independent and \( 
	\Theta_1,\ldots,\Theta_I \) are identically distributed.
	
	\item Random variables $\mu(\Theta_i)$ and $\sigma^2(\Theta_i)$ possess the following 
	finite moments for all $i \in \{1,2,\dots,I\}$:
	\begin{enumerate}
		\item[] \( \boldsymbol{\mu}_0 = \mathbb{E}[\boldsymbol{\mu}(\Theta_i)] \), 
		\item[] \( \boldsymbol{\sigma} = 
		\mathbb{E}[\boldsymbol{w}_{ti}\mathbb{E}[\mathrm{Cov}(\mathbf{X}_{ti}, 
		\mathbf{X}_{ti}^{'}  \mid \Theta_i)]\boldsymbol{w}_{ti}] \), and
		\item[] \( \boldsymbol{\tau} = \mathrm{Cov}[\boldsymbol{\mu}(\Theta_i), 
		\boldsymbol{\mu}(\Theta_i)^{'}] \), 
	\end{enumerate}
	where $\boldsymbol{\mu}_0$, $\boldsymbol{\sigma}$ and $\boldsymbol{\tau}$ represent the 
	global means, the within-group covariance matrix and the between-group covariance matrix 
	respectively.
\end{enumerate}

The multidimensional credibility estimator is a linear combination of the individual experience vector
\(
\mathbf{\overline{X}}_{i} = (\overline{X}_{i}^{(1)}, \dots, \overline{X}_{i}^{(M)})
\), which are the condensed random variables defined as
\begin{equation*}
	\overline{X}_i^{(m)} = \sum_{t=1}^{T} \frac{X_{ti}^{(m)} w_{ti}^{(m)}}{\sum_{s=1}^{T} 
		w_{si}^{(m)}},
\end{equation*}
and the collective mean
\(
\widehat{\boldsymbol{\mu}_{0}} = (\sum_{i=1}^{I} \boldsymbol{\alpha}_{i})^{-1} \sum_{i=1}^{I} 
\boldsymbol{\alpha}_{i} \mathbf{\overline{X}}_{i}
\).
This is mediated by the credibility weights matrix
\begin{equation*}
\boldsymbol{\alpha}_{i} = \boldsymbol{\tau}(\boldsymbol{\tau} + \boldsymbol{w}_{\bullet i}^{-1/2} 
\boldsymbol{\sigma} \boldsymbol{w}_{\bullet i}^{-1/2})^{-1}
\end{equation*}
and yields
\begin{equation}
	\label{eq:cred_estimators_mult}
\widehat{\boldsymbol{\mu}(\Theta_i)} = \boldsymbol{\alpha}_{i} \mathbf{\overline{X}}_{i} + (I 
- \boldsymbol{\alpha}_{i}) \widehat{\boldsymbol{\mu}_{0}}.
\end{equation}

\vspace{1cm}

\noindent \textbf{Remarks}: 
\begin{remark}
	The higher the diagonal elements of the credibility weight matrix 
	$\boldsymbol{\alpha}_i$, the more the condensed claims experience 
	$\overline{X}_i$ is weighted when calculating the credibility estimator. High credibility 
	weights are observed in risks with high volume measures $w_{ti}$ and are generally higher 
	when the within-group variance is low and the between-group variance  
	is high.
\end{remark} 

\begin{remark} 
	\label{rem:estimation} 
	Estimators for $\boldsymbol{\sigma}$ and $\boldsymbol{\tau}$ are discussed in 
	\cite{buhlmann2005course}. 
\end{remark} 

\begin{remark} 
	\label{rem:tau_adjustment} 
	In practice, the estimated covariance matrix $\boldsymbol{\tau}$ may contain off-diagonal 
	elements that are excessively large relative to the corresponding diagonal elements. For 
	$M=2$, this occurs when \[ \tau_{12}\tau_{21} > \tau_1^2\tau_2^2. \] Bühlmann and Gisler 
	\cite{buhlmann2005course} 
	suggest reducing the off-diagonal elements such that \[ \tau_{12}\tau_{21} \leq 
	\tau_1^2\tau_2^2. \] However, this adjustment may render $\boldsymbol{\tau}$ singular. In 
	practice, if $\boldsymbol{\tau}$ is singular, a shrinkage factor $\rho \in [0,1]$ is 
	introduced and multiplied with the off-diagonal elements. 
\end{remark} 

\begin{remark} 
	\label{rem:rho_zero} 
	The choice $\rho=0$ yields the classical Bühlmann--Straub solution 
	\cite{buehlmann_straub}. 
\end{remark}

\subsection{Clustering}
\label{sec:clustering}
Clustering refers to a class of unsupervised learning methods that partition unlabeled data 
into groups based on a defined notion of similarity. Examples of actuarial applications can 
be found in \cite{Gan02042020}. Common approaches include partitioning methods such as 
k-means \cite{k-means,k-means_2}, hierarchical clustering \cite{ward1963hierarchical}, and 
density-based methods such as DBSCAN \cite{dbscan}. 
\par
In this paper, the objective is to obtain an optimal cluster allocation for a predetermined 
number of risk groups. We therefore focus on partitioning methods, specifically k-means, 
which directly optimize a clustering objective function. Alternative clustering approaches 
may be investigated in future research.
\\ \noindent 
\\ \noindent
\textbf{K-means}
\\ \noindent
K-means is a partitioning clustering method that minimizes the total within-cluster sum of 
squared Euclidean distances between data points and their respective cluster centers.
Let $\mathcal{C}=\{\mathcal{C}_1,\dots,\mathcal{C}_K\}$ denote a partition of the 
observations into $K$ clusters.
Furthermore, let
\begin{equation*}
	N_k = \sum_{i \in \mathcal{C}_k} 1
\end{equation*}
denote the number of data points in cluster $\mathcal{C}_k$ and
\begin{equation}
	\label{eq:clus_alloc}
	\overline{\boldsymbol{\mu}}_{k} :=  
	(\overline{\mu}_{k}^{(1)},...,\overline{\mu}_{k}^{(M)})^{T}   \text{,  where }   
	\overline{\mu}_{k}^{(i)} = \frac{1}{N_k} \sum_{n \in \mathcal{C}_{k}} \mu_{n}^{(i)}
\end{equation}
be the vector of component-wise means over the $M$-dimensional data points in cluster 
$\mathcal{C}_k$. A cluster allocation is obtained by assigning data points to clusters, such 
that the total within-cluster variance is minimized
\begin{equation}
	\label{eq:clus_min}
	\mathcal{C}^{*} = \arg \min_{\mathcal{C}} \sum_{k=1}^{K} \sum_{i \in \mathcal{C}_{k}} || 
	\boldsymbol{\mu}_i - \overline{\boldsymbol{\mu}}_{k} ||^{2},
\end{equation}
where $|| \cdot ||$ represents the Euclidean distance.
The optimization problem in Equation~\eqref{eq:clus_min} is solved iteratively by alternating 
between assigning observations to their nearest cluster center and updating the cluster 
centers according to Equation~\eqref{eq:clus_alloc} until convergence.

\subsection{The Credibility Clustering Framework (CCF)}
\label{sec:cred_clustering}
\begin{figure}[H]
	\centering
	\includegraphics[width=\textwidth]{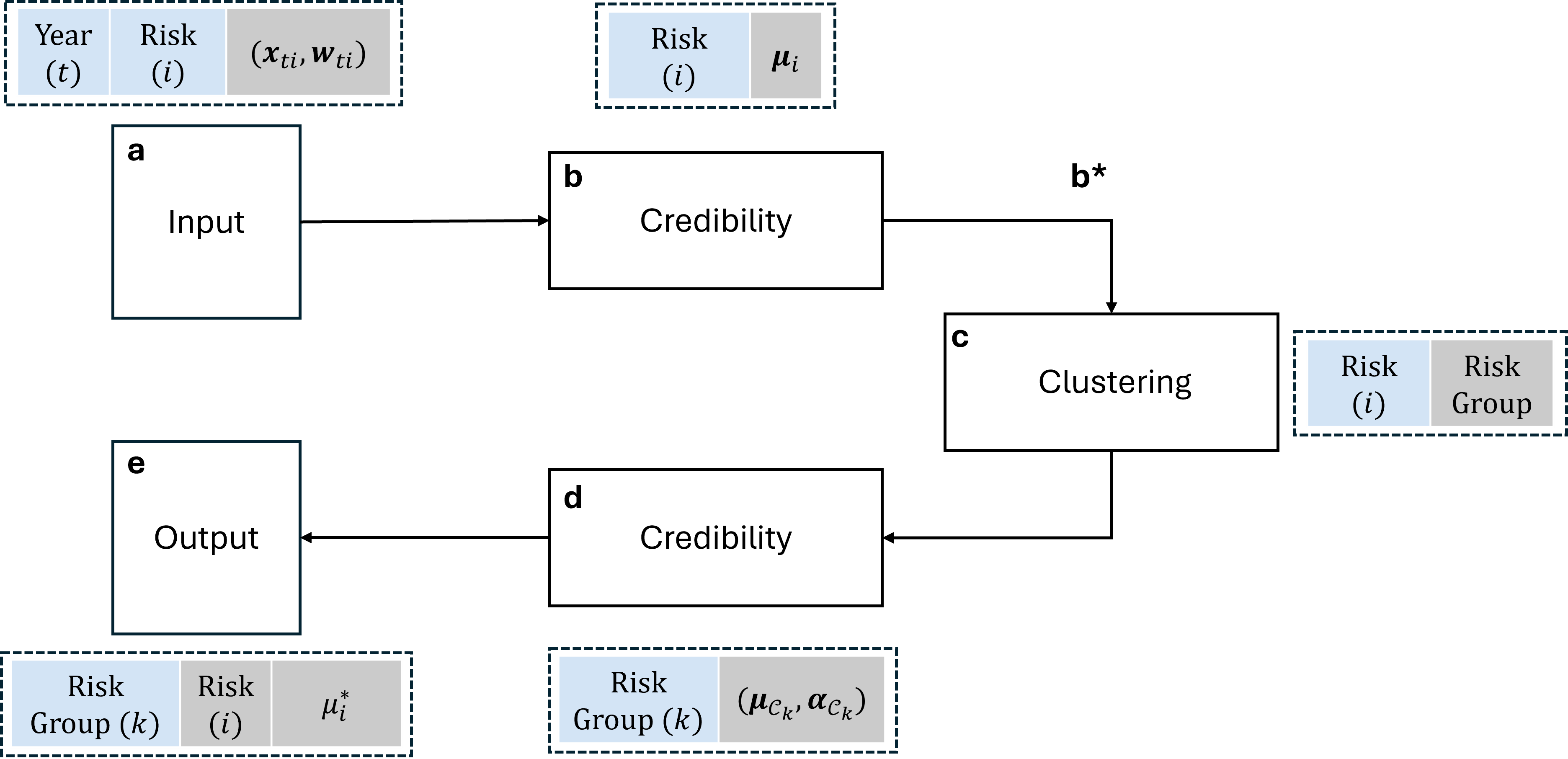}
	\caption{Schematic overview of the credibility clustering framework. Squares denote input and output stages; rectangles represent the methodological components (credibility and clustering). In our application, we use the multidimensional B\"uhlmann--Straub model and k-means. Gray and blue cells specify relevant variables and granularity levels.}
	\label{cred_clus_diag}
\end{figure}

Credibility theory and clustering are combined to form the credibility clustering framework (CCF) (cf. Figure~\ref{cred_clus_diag}). The individual steps of the framework are explained in more detail below.
\begin{enumerate}
	\item[(\textbf{a})] The framework requires claims experience -- as defined by $(\boldsymbol{x}_{ti},\boldsymbol{w}_{ti}) \in \mathbb{R}^{M} \times \mathbb{R}^{M}$ -- per risk $i$ and year $t$ (cf. Table~\ref{table:setting}).
	\item[(\textbf{b})] The credibility model -- in our case the multidimensional Bühlmann-Straub model (cf. Subsection~\ref{sec:mbs}) -- is applied to all $M$ risk measures simultaneously. This condenses the claims experience $(\boldsymbol{x}_{ti}, \boldsymbol{w}_{ti})$ into a vector of credibility estimates $\boldsymbol{\mu}_{i} \in \mathbb{R}^{M}$ (cf. Equation~\eqref{eq:cred_estimators_mult}). The resulting estimates incorporate the statistical reliability of the observations, which are captured by the diagonal elements of the credibility weight matrix $\boldsymbol{\alpha}_{i}$.
	\item[($\mathbf{b^{*}}$)] Optional for high-dimensional data ($M \geq 3$): A dimensionality reduction technique -- for example Principal Component Analysis~\cite{PCA_1} \cite{hotelling1933analysis} -- can be utilized to project the $M$-dimensional vector of credibility estimators onto a lower, $D$-dimensional representation ($\mathbf{z}_{i} \in \mathbb{R}^{D}$) before applying the clustering. 
	\item[(\textbf{c})] Clustering -- in our use case k-means -- is applied to the credibilized data to group the $I$ risks into $K$ risk groups, resulting in risk groups ($\mathcal{C}_{1}, \dots, \mathcal{C}_{K}$). As k-means is initialized randomly and thereby sensitive to variation in solutions depending on the starting position, 100 unique starting positions are  randomly initiated, after which the best solution -- as measured by the within-variance -- is selected. Afterwards, the risk groups are ordered according to the means of their respective cluster centers $\mathbf{c}_{k}$, such that the first risk group represents the risks with the lowest credibilized claims experience and the $K$-th risk group represents the risks with the highest credibilized claims experience.
	\item[(\textbf{d})] The same credibility model from Step (\textbf{b}) is applied to all risk measures of the original input data (\textbf{a}), which is now aggregated to a risk group $k$ per year $t$ level.
	This yields credibility estimates for all $K$ risk groups $\boldsymbol{\mu}_{\mathcal{C}} = (\boldsymbol{\mu}_{\mathcal{C}_{1}}, \dots, \boldsymbol{\mu}_{\mathcal{C}_{K}}) \in \mathbb{R}^{K \times M}$, which are referred to as \textit{risk group factors}.
	\item[(\textbf{e})] Finally, the risk group factors for all risk measures are combined into a single factor by taking the weighted sum with the diagonal elements of the credibility weights matrix $\boldsymbol{\alpha}_{\mathcal{C}_k}$ (cf. Remark~\ref{rem:waagschaal})
	\begin{equation}
		\label{eq:waagschaal}
		\mu_{\mathcal{C}_k}^{*} = \sum_{m=1}^{M} \mu_{\mathcal{C}_k}^{(m)^{*}} \mfrac{ \alpha_{\mathcal{C}_k}^{(m)}}{\sum_{r=1}^{M} \alpha_{\mathcal{C}_k}^{(r)}}.
	\end{equation}
	The resulting combined risk group factors are then assigned to individual risks according to the clustering solution obtained in Step (\textbf{c}). After this assignment, these are referred to as \textit{tariff factors} and are denoted by $\mu_{i}^{*}$.
\end{enumerate}
\noindent
\textbf{Remarks:}
\begin{remark}
	\label{rem:waagschaal}
We use the diagonal elements as an interpretable proxy for the direct credibility contribution of each risk 
measure. This aggregation rule is a pragmatic choice and is not implied by the optimality properties of the 
multidimensional Bühlmann–Straub estimator.
\end{remark}

\section{Results}
\label{sec:results}
\begin{figure}[H]
	\centering
	\includegraphics[width=\textwidth]{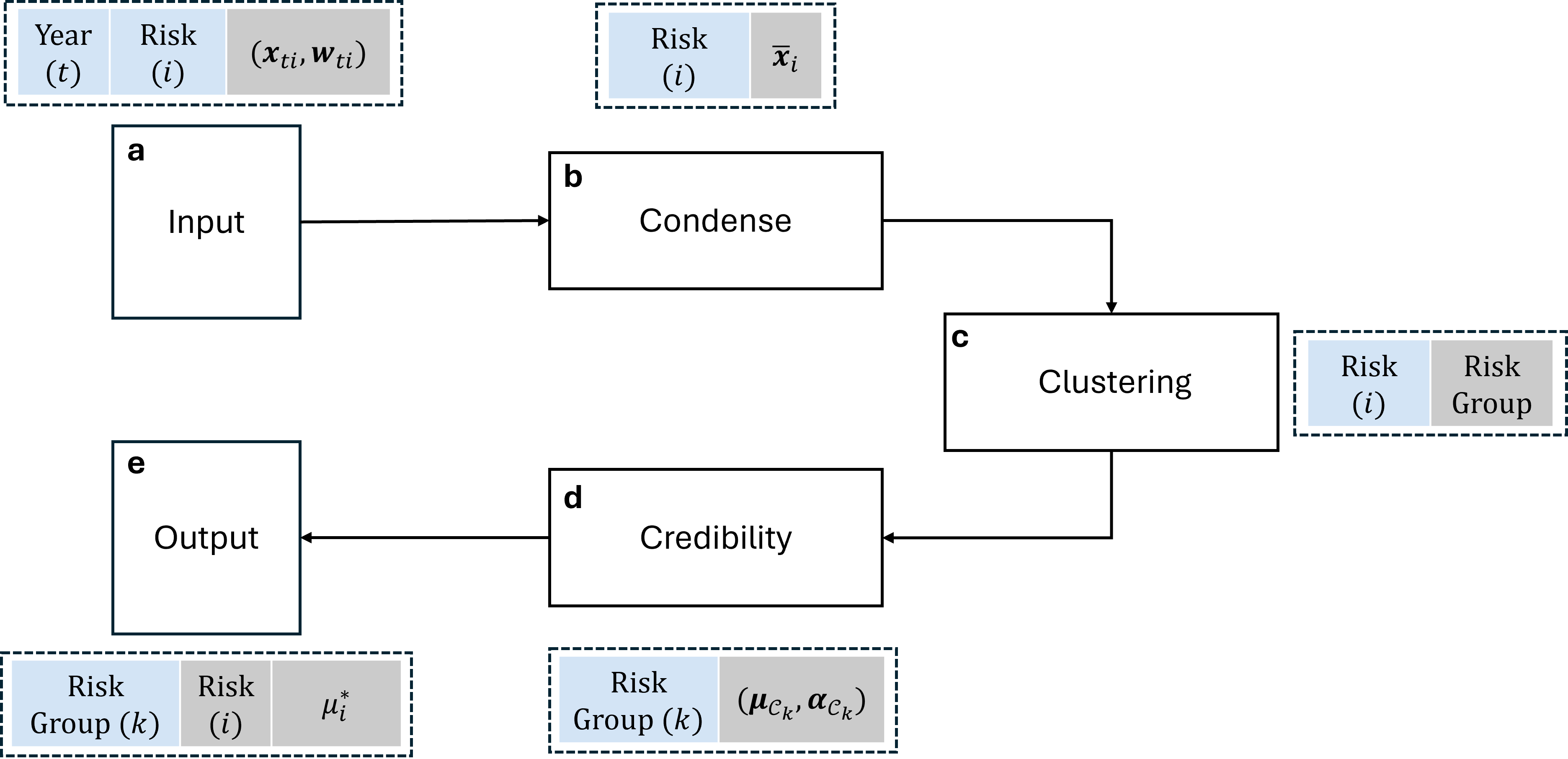}
	\caption{Overview of the k-means method, with and without outlier removal, used as a benchmark for the CCF. In contrast to credibility clustering, the input data is directly aggregated using Equation~\eqref{eq:condensed_statistic}. Depending on the variant, outliers are either retained or removed from the data. Clustering is then applied to this aggregated data. If outliers were removed beforehand, they are subsequently reassigned to the nearest cluster center. Finally, the credibility model is applied to the resulting clusters (i.e. risk groups).}
	\label{fig:comp_meth}
\end{figure}

The quality of the credibility clustering framework (CCF) is assessed
based on its performance with respect to the three key methodological and practical challenges outlined in the introduction:
\begin{enumerate}
	\item Differentiation of observations according to their statistical reliability, 
	\item Practical viability of the resulting clusters,
	\item Temporal stability of the clustering solution.
\end{enumerate}
To clarify how performance is evaluated, we first describe the two underlying data sources. Subsequently, the results for each challenge are discussed and benchmarked against two alternative approaches: direct application of k-means to the claims experience with and without outlier exclusion (cf.  Remark~\ref{rem:outlier} and Figure~\ref{fig:comp_meth}) based on the condensed claims experience
\begin{equation}
	\label{eq:condensed_statistic}
	\overline{x}_i^{(m)} = \sum_{t=1}^{T} \frac{x_{ti}^{(m)} w_{ti}^{(m)}}{\sum_{s=1}^{T} 
		w_{si}^{(m)}}.
\end{equation}
.
\noindent
\textbf{Remarks:}
\begin{remark}
	\label{rem:outlier}
	Outliers are defined as risks for which the claims experience of one or more risk measures deviates from 
	the corresponding mean by more than two standard deviations.
\end{remark}

\subsection{Data}
\label{subsec:data}
The empirical performance of the CCF is evaluated using mortality data from a large European life insurance company (cf. Tables~\ref{tab:data} and \ref{tab:ec_sec_data}). The data are aggregated at the year-economic sector $(t,i)$ level, where economic sectors group insured companies according to their underlying economic activity. Table~\ref{tab:ec_sec_data} reveals substantial heterogeneity in the amount of available data across sector-year observations.
\begin{table}[htbp] 
	\centering   
	\begin{tabular}{p{1.5cm}|p{2cm}|p{1.5cm}|p{2cm}|p{3cm}}
		 \toprule 
		 \textbf{Years} & \textbf{Contracts} & \textbf{Insured} & \textbf{Economic} & \textbf{Risk} \\ $t$ & & & \textbf{Sectors} $i$ & \textbf{Measures} $m$ \\ 
		 \midrule 
		 2014-- 2022 & 36\,669-- 42\,624 & 327\,478-- 397\,216 & 84 & \{Claims Frequency, Claims Severity\} \\ \bottomrule 
\end{tabular} 
\caption{Overview of the available data. The numbers of contracts and insured individuals are presented as ranges across the observation period. The lower value corresponds to the minimum annual count, while the upper value corresponds to the maximum annual count.}
\label{tab:data}
\end{table}
\begin{table}[H] 
\centering  
	\begin{tabular}{p{1.5cm}|p{1.4cm}|p{2.5cm}|p{2.5cm}|p{2cm}} 
		\toprule \textbf{Statistic} & \textbf{No.\newline Insured} & \textbf{No. Expected Cases} & \textbf{Insured Amount} & \textbf{Expected Claim} \\ 
		\midrule \textbf{Min} & 1 & approx. 1/1000 & 6\,449 CHF & 6 CHF \\ 
		\textbf{Q25} & 602 & 0.61 & 6.7 Mio. CHF & 7\,109 CHF \\ 
		\textbf{Median} & 1\,743 & 1.79 & 24.6 Mio. CHF & 26\,102 CHF \\ 
		\textbf{Q75} & 4\,833 & 5.32 & 60.9 Mio. CHF & 71\,115 CHF \\ 
		\textbf{Max} & 37\,648 & 40.64 & 803.1 Mio. CHF & 935\,704 CHF \\ 
		\bottomrule 
	\end{tabular}
\caption{Summary statistics of the data at the year-economic sector $(t,i)$ level.}
\label{tab:ec_sec_data}  
\end{table}
At the year-economic sector granularity, there are two available risk measures:
\begin{enumerate}
	\item \textbf{Claim Frequency}: The volume measure $w_{ti}^{(1)}$ is the expected number of claims for sector $i$ in year $t$ (cf. Remark~\ref{rem:exp}). The claims experience $x_{ti}^{(1)}$ is the ratio between the observed and the expected number of claims.
	\item \textbf{Claim Severity}:  The volume measure $w_{ti}^{(2)}$ is the expected annual survivor's pension for sector $i$ in year $t$ (cf. Remark~\ref{rem:exp}). The claims experience $x_{ti}^{(2)}$ is the ratio between the observed and the expected annual survivor's pension.
\end{enumerate}
To measure the temporal stability of the framework as more data is available, the CCF is performed on an incrementally growing set of years. The first set of years contains $S_{1} := \{2014,\dots,2017\}$, which represents the data used to price contracts in year $2019$ (cf. Remark~\ref{rem:eingabe}). Set one is hence called \textit{tariff year} 2019. Tariff year 2020 (i.e. $S_{2}$) additionally contains year $2018$ and so on (cf. Remark~\ref{rem:ibnr}).
\par
Given the available number of years $T$, there are six available sets of years 
$\{S_{1}, \dots, S_{6}\}$ and five observable transitions between sets of years. 
This provides a basis for assessing whether the statistical reliability is 
appropriately accounted for by the CCF. However, the five transitions do not 
provide enough temporal depth to evaluate more data-intensive statistical 
characteristics, such as distributional properties. To address this limitation, 
we apply bootstrapping to set $S_{6}$. 

\vspace{0.5cm} \noindent  
\textbf{Bootstrapping}
\noindent \\
Bootstrapping is a statistical method which uses resampling with replacement on observations of a dataset to allow for statistical evaluation without making assumptions about the underlying distribution \cite{Efron1992}. Bootstrapping has already been applied in the actuarial field, examples include \cite{ENGLAND1999281}. For a more detailed explanation of bootstrapping in the actuarial context, we refer to \cite{wüthrich2008stochastic}.
\par 
In the context of this paper, we use bootstrapping to resample (with replacement) the data for each unique combination of an economic sector and an observation year for the available years $t \in \{2014, \dots, 2022\}$. For each combination $(t, i)$, we have a list of contracts assigned to the respective economic sector, each with an observed- and expected claim experience associated with that observation year. The bootstrap procedure resamples this list of contracts and their associated claims experience, which, when aggregated, leads to a modified claims experience for the corresponding economic sector and year. 
\par
The tariff year 2025 ($S_6$) dataset is bootstrapped 1\,000 times, generating 1\,000 bootstrap samples. For each of these bootstrap samples, the CCF and the k-means approaches (both with and without outliers) are applied. The bootstrap analysis serves as a perturbation-based robustness assessment complementing the observed tariff-year transitions.

\vspace{0.5cm} 
\noindent 
\textbf{Parameters} 
\noindent \\ 
The raw estimates of $\boldsymbol{\tau}$ are non-viable for the available mortality data (cf. 
Remark~\ref{rem:tau_adjustment}). 
Therefore, a shrinkage factor of $\rho = 0.35$, calibrated according to the procedure described in 
Remark~\ref{rem:rho_mort}, is applied to the off-diagonal elements of $\boldsymbol{\tau}$. This calibration 
is based on the 
assumption that the credibility-induced increase in dependence between the risk measures is comparable for 
mortality and disability experience. A sensitivity analysis 
was performed to assess the impact of different values of 
$\rho$ on the results (cf. Appendix 
Section~\ref{sec:app_rho}). The results are relatively insensitive to moderate and large values of $\rho$. 
The proportion of changes exceeding 10\% decreases substantially between $\rho=0$ and $\rho=0.35$, whereas 
no systematic further improvement is observed for larger values.
\par 
Furthermore, the clustering procedure is performed using five cluster centers ($K=5$). This number was deemed sufficient 
to capture the heterogeneity in the data while maintaining the statistical reliability of the resulting 
clusters. A sensitivity analysis was performed for three 
($K=3$) and seven cluster centers ($K=7$) (cf. Appendix 
Section~\ref{sec:app_clus}). Although other odd numbers of 
cluster centers could equally be considered, the impact of 
this modeling choice on the proposed framework is beyond 
the scope of the present paper. The qualitative conclusions remain largely robust across $K=3,5,7$, although 
the relative advantage of CCF becomes less uniform for $K=7$, particularly for tariff-year stability at the 
sector level.
\par
To mitigate the risk of convergence to a poor local optimum, the clustering algorithm is initialized from 100 
randomly selected starting configurations, and the solution with the smallest within-cluster variance is 
retained. Given the relatively small number of observations and dimensions considered, many of the random 
initializations converge to the same clustering solution, indicating a high degree of stability of the selected 
partition.

\vspace{0.5cm}
\noindent \textbf{Remarks}
\begin{remark}
	\label{rem:exp}
	The expected number of claims and the expected annual survivor's pension are calculated based on the mortality tables from the insurance company.
\end{remark}
\begin{remark}
	\label{rem:eingabe}
	The gap of one year is due to the yearly quotation season starting a year prior to the start of the insurance coverage.
\end{remark}
\begin{remark}
	\label{rem:ibnr}
	The reporting date of claims is also taken into account: Claims that are reported in the span of 2018 but relate to cases in 2014 until 2017 are also updated in future sets $S$.
\end{remark}
\begin{remark}
	\label{rem:rho_mort}
	The shrinkage factor is estimated by matching the post-credibility correlation structure of the mortality data to that observed for the disability data. For the disability portfolio, the multidimensional credibility model provides viable estimates of the covariance matrix $\boldsymbol{\tau}$, yielding reliable post-credibility correlations between the risk measures. We assume that the proportional increase in correlation induced by the multidimensional credibility approach, relative to the one-dimensional Bühlmann-Straub credibility models, is similar for mortality and disability experience. The shrinkage factor is then obtained by solving for the value that reproduces the corresponding post-credibility correlation level in the mortality portfolio.
\end{remark}

\subsection{Statistical Reliability}
\label{sec:res_ss}
Accounting for the statistical reliability of the observed claim experience is the primary objective of the CCF. This means that the more statistically significant an observed claims experience is, the more it influences the assigned risk group factors and vice versa. However, we must first define how statistical reliability is measured.
\par 
The statistical reliability of the observed claims experience is quantified through the diagonal elements of the credibility matrix $\boldsymbol{\alpha}_{i}$. To obtain a one-dimensional reliability indicator, the mean of the two diagonal elements of the credibility matrix is calculated and subsequently normalized to the interval $[0,1]$.
\par 
To analyze the effects of credibility clustering, we first examine its impact on the claims experience. Subsequently, we assess how it affects the distribution of statistically reliable economic sectors across the resulting risk groups. 
\par
\begin{figure}[H]
	\centering
	\includegraphics[width=\textwidth]{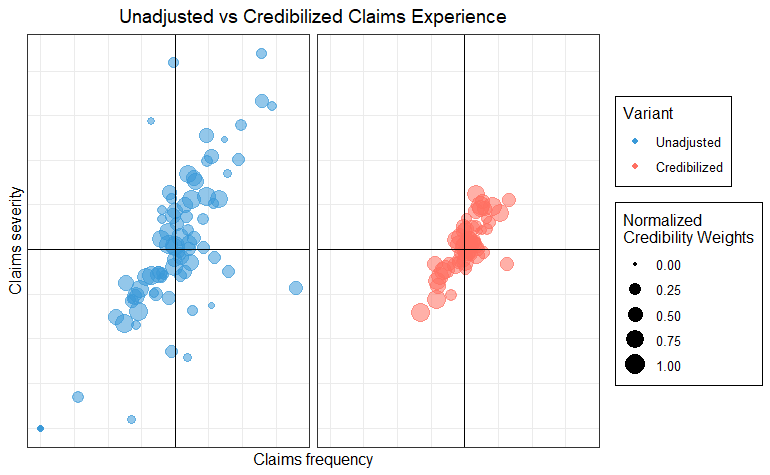}
	\caption{Comparison of the unadjusted (blue, left) and the credibilized claims experience (red, right). The size of each point reflects the proxy for statistical reliability, defined as the normalized credibility weight. Larger points correspond to higher statistical reliability. The intersection of the vertical and horizontal black lines indicates the overall portfolio experience at (1,1).}
	\label{fig:stat_sig_ote}
	
\end{figure}

 
Figure~\ref{fig:stat_sig_ote} shows the impact of applying the B\"uhlmann-Straub model to the claims experience. The unadjusted data (blue, on the left) includes many smaller economic sectors located at the extremities of the plot. Economic sectors without any observed mortality cases are clustered at (0,0). In contrast, small economic sectors with one or more observed cases tend to exhibit a claims experience substantially greater than 1, as their expected number of cases is significantly below 1. 
\par
The credibilized claims experience shows the opposite pattern. The outer edges of the plot are now populated by the larger economic sectors, while the smaller economic sectors are located between the extremities and the center, defined by the intersection of the black lines at (1,1).
\par 
Calculating the empirical Pearson correlation coefficient between the unadjusted or credibilized claims experience and the corresponding credibility weights for each risk measure confirms this effect (cf. Table~\ref{tab:cor_ote_vol}). The unadjusted experience exhibits a negative correlation with the credibility weights, indicating that higher observed-to-expected ratios tend to be associated with lower credibility weights. In contrast, the credibilized claims experience shows a positive correlation with the credibility weights.
\begin{table}[htbp]
	\centering
	\begin{tabular}{lrr}
		\toprule
		\textbf{Claims Experience} & \textbf{Frequency} & \textbf{Severity} \\
		\midrule
		Credibilized & 38.8\% & 49.4\% \\
		Unadjusted   & -44.4\% & -47.0\% \\
		\bottomrule
	\end{tabular}
	\caption{Empirical Pearson correlation between the credibilized and unadjusted claims experience and the credibility weights of the NOGA-codes.}
	\label{tab:cor_ote_vol}
\end{table}
\par
\begin{figure}[H]
	\centering
	\includegraphics[width=\textwidth]{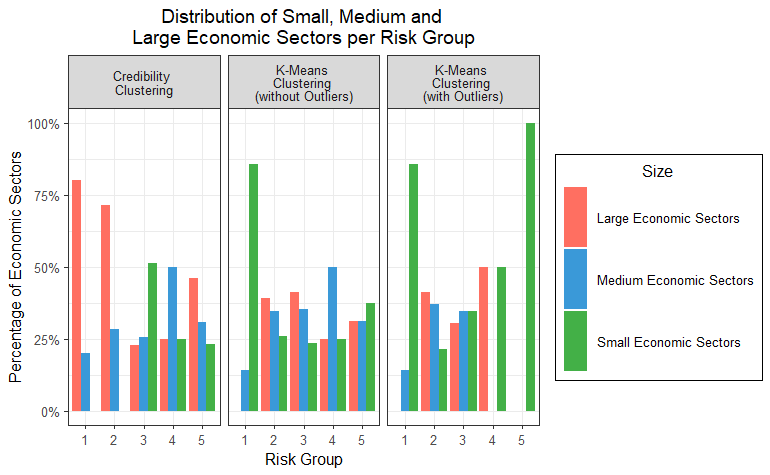}
	\caption{Figure showing the distribution of small (green), medium (blue) and large (red) economic sectors within the risk groups depending on the type of clustering used. Small, medium and large economic sectors correspond to the bottom, middle and top 33.3\% of economic sectors, as measured by their summed normalized volumes.}
	\label{fig:stat_sig_rk}
\end{figure}

The difference in the economic sectors occupying the extremes of the distribution also affects the composition of the risk groups (cf. Figure~\ref{fig:stat_sig_rk}). When the Bühlmann-Straub model is applied beforehand, the extreme risk groups (in this example, groups one and five, representing the best and worst risks, respectively) are primarily populated by large economic sectors (red). In contrast, without prior credibility adjustment, these groups consist predominantly of medium-sized (blue) and small (green) economic sectors.
\par
All in all, applying the B\"uhlmann-Straub model prior to clustering results in a more equitable distribution of economic sectors across the risk groups. More specifically, it enables rewarding (or penalizing) economic sectors for their individual claims experience relative to the overall collective, while appropriately accounting for their statistical reliability. 

\subsection{Viability}
\label{sec:res_v}
\begin{figure}[H]
	\centering
	\includegraphics[width=\textwidth]{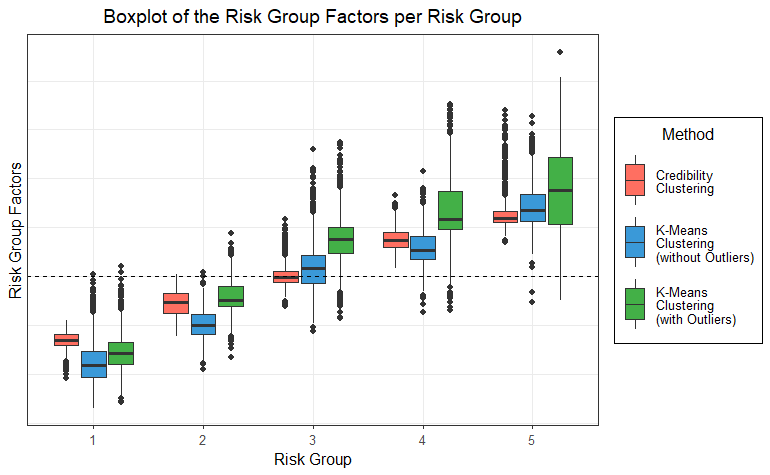}
	\caption{Boxplots of the risk group factors for risk groups one to five, comparing the CCF (red) with k-means clustering, both excluding (blue) and including (green) outliers. The boxes represent the interquartile range (IQR), with whiskers extending up to 1.5 times the IQR. Dots indicate outliers beyond this range. The dashed horizontal line indicates the global risk experience (value of 1). The y-axis is omitted for confidentiality reasons.}
	\label{fig:RK_viability}
\end{figure}

Accounting for the statistical reliability of observations alone is not sufficient for implementing the CCF in practice. In addition, the resulting classification must be viable. We assess the viability of the solution based on the following criteria:
\begin{enumerate}
	\item Risk-group differentiation: Risk group factors should increase monotonically across risk groups, remain sufficiently differentiated, and exhibit approximate symmetry around the central risk group (in this case risk group three).
	\item Central alignment: The risk group factor of the middle risk group should be close to the global average risk experience (in our use case close to 1).
	\item Temporal stability: The resulting tariff factors should stay stable across time (i.e. tariff year changes), unless there are significant changes in the observed claims experience.
\end{enumerate}
To evaluate the CCF with respect to these criteria, we apply bootstrapping as described at the beginning of Section~\ref{sec:results}. The results are shown in Figure~\ref{fig:RK_viability}, which addresses the first two viability criteria listed above. The third criterion -- temporal stability -- is discussed separately in Section~\ref{sec:res_ts}. 
Figure~\ref{fig:RK_viability} shows that the risk group factors obtained from 
k-means clustering exhibit substantial variability, leading to violations of the 
risk-group differentiation criterion. In particular, for k-means clustering 
including outliers, the range of risk group factors for risk group five partially falls 
below the global experience level. Furthermore, for k-means clustering excluding 
outliers, the ranges of risk group factors for risk groups three and four overlap 
substantially. Among the three approaches, the CCF produces the most 
differentiated risk group structure.
\par 
The middle risk group, in this case risk group three, is appropriately placed near the global experience of one for the CCF. Additionally, the variance is low, indicating that this is consistently the case. Similarly, k-means clustering without outliers shows a well-placed median; however, the variance is large, implying that this placement is unstable. K-means clustering with outliers places the risk group factor for risk group three significantly too high and is therefore not viable.
\par 
Overall, the CCF satisfies the first two evaluation criteria and outperforms the 
benchmark approaches in this use case. The third criterion, temporal stability, 
is examined below.

\subsection{Temporal Stability}
\label{sec:res_ts}
The temporal stability of the CCF is evaluated by analyzing both the tariff year data and the bootstrap data. In the evaluation of the tariff year data, we focus on the \textbf{risk group stability} and \textbf{tariff factor stability} as tariff years change. These elements cannot be evaluated for the bootstrapped data, because bootstrap samples are inherently not ordered along a temporal dimension. Furthermore, we analyze the standard deviation of tariff factors and risk group factors for both data sets.

Below, we start by analyzing the changes in tariff factor, followed by an examination of the risk group stability and standard deviation as potential drivers for these changes.

\noindent \\
\textbf{Absolute tariff factors changes}
\noindent \\
To assess temporal stability, we analyze the absolute tariff factor changes 
\begin{equation} 
	\label{delta} 
	\Delta^{(S_{s+1})}_i = \left| \frac{\mu_i^{*(S_{s+1})}} {\mu_i^{*(S_s)}} - 1 \right|, \quad s \in \{1,\dots,5\}, 
\end{equation}
where $\mu_i^{*(S_{s})}$ denotes the tariff factor for economic sector $i$ obtained from either the CCF (cf. Figure \ref{cred_clus_diag}) or a benchmark method (cf. Figure \ref{fig:comp_meth}) using the set of years $S_{s}$ (cf. Subsection~\ref{subsec:data}). Building on these absolute changes, two temporal stability measures are considered:
\begin{itemize}
\item \textbf{Average Absolute Change}: The mean across all tariff year changes based on Equation \eqref{delta}.
\item \textbf{Proportion of Changes $\geq$ 10\%}: The proportion of absolute changes that exceed 10\% across all tariff year changes based on Equation \eqref{delta}.
\end{itemize}

Table~\ref{tab:c5_method_comparison} summarizes the results. The CCF yields the lowest average absolute 
change of 7.21\%, compared with 16.60\% for k-means without outliers and 14.46\% for k-means including 
outliers. A similar pattern is observed for the proportion of changes exceeding 10\%, where the CCF 
achieves a value of 23.57\%, substantially below the corresponding values of 60.71\% and 52.62\%.
\par 
The 10\% threshold is chosen as a practically relevant benchmark, as tariff factor changes of this 
magnitude are commonly regarded as material in experience-rating applications. The CCF yields the 
lowest proportion of changes exceeding this threshold.
\par
\begin{figure}[h]
	\centering
	\includegraphics[width=\textwidth]{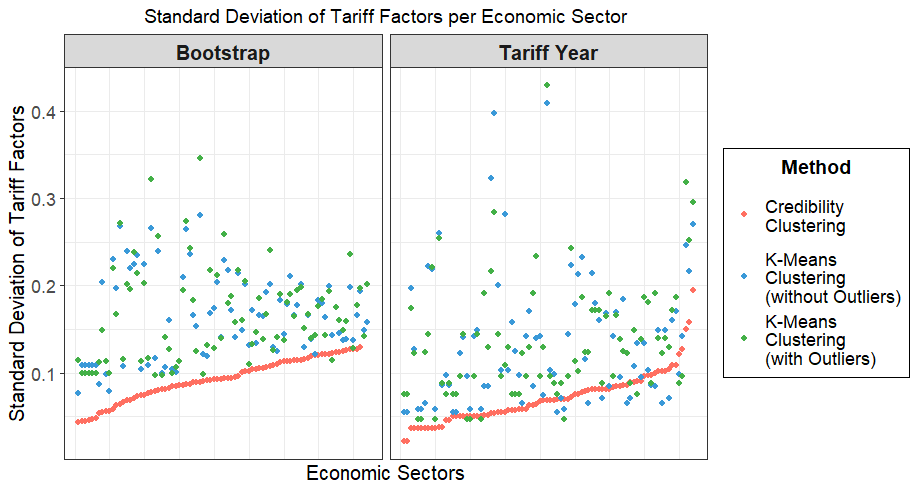}
	\caption{Graph comparing the standard deviation of the tariff factors across bootstrap samples and tariff year changes per economic sector and method. The data are ordered by the standard deviation of the CCF risk group factors. The x-axis labels are omitted due to confidentiality.}
	\label{fig:SD_EcSec}
\end{figure}

Overall, the CCF delivers the most stable results among the three methods. Nevertheless, both the 
average tariff factor changes and the frequency of changes exceeding 10\% remain substantial. The 
causes of this residual volatility are therefore investigated further below.
\begin{table}[H]
	\centering
	\small
	\begin{tabular}{lcc}
		\hline
		\multicolumn{1}{l}{\textbf{Method}} &
		\multicolumn{1}{c}{\makecell{\textbf{Average Absolute} \\ \textbf{Change}}} &
		\multicolumn{1}{c}{\makecell{\textbf{Proportion of Changes} \\ \textbf{$\geq$ 10\%}}} \\
		\hline
		CCF & 7.21\% & 23.57\% \\
		K-Means (without Outliers)& 16.60\% & 60.71\% \\
		K-Means (with Outliers) & 14.46\% & 52.62\% \\
		\hline
	\end{tabular}
	\caption{Average absolute change and proportion of tariff-factor changes exceeding 10\% across all tariff-year changes for the three methods.}
	\label{tab:c5_method_comparison}
\end{table}
\noindent
\textbf{Standard deviation}
\noindent \\
The volatility of the tariff factors across tariff year changes is influenced by the volatility of the risk group factors and the risk group allocation of the economic sectors. Tables~\ref{tab:c5_sd_rg_tariff_year} and \ref{tab:c5_sd_rg_bootstrap} show the standard deviation of the risk group factors. The relative differences between the methods are substantial, ranging from approximately a factor of two to five. The CCF produces the most stable risk group factors in all risk groups for the tariff year data and the bootstrap samples. The results are similar for the standard deviation of the tariff factors across economic sectors (cf. Figure~\ref{fig:SD_EcSec}).

 \begin{table}[H] 
 	\centering 
 	\begin{tabular}{lccccc} 
 		\toprule & \multicolumn{5}{c}{\textbf{Risk Group}} \\ 
 		\cmidrule(lr){2-6} \textbf{Method} & \textbf{1} & \textbf{2} & \textbf{3} & \textbf{4} & \textbf{5} \\ 
 		\midrule Credibility Clustering & 0.017 & 0.042 & 0.035 & 0.034 & 0.076 \\ 
 		K-Means (without Outliers) & 0.055 & 0.049 & 0.075 & 0.081 & 0.158 \\ 
 		K-Means (with Outliers) & 0.045 & 0.061 & 0.088 & 0.168 & 0.196 \\ 
 		\bottomrule 
\end{tabular} 
\caption{Standard deviation of risk group factors across tariff years by risk group and method.} 
\label{tab:c5_sd_rg_tariff_year} 
\end{table} 
 
 \begin{table}[H] 
 	\centering 
 	\begin{tabular}{lccccc} \toprule & \multicolumn{5}{c}{\textbf{Risk Group}} \\
 		 \cmidrule(lr){2-6} \textbf{Method} & \textbf{1} & \textbf{2} & \textbf{3} & \textbf{4} & \textbf{5} \\ 
 		 \midrule Credibility Clustering & 0.040 & 0.064 & 0.055 & 0.056 & 0.062 \\ 
 		 K-Means (without Outliers) & 0.102 & 0.071 & 0.124 & 0.082 & 0.106 \\ 
 		 K-Means (with Outliers) & 0.091 & 0.092 & 0.113 & 0.156 & 0.230 \\ 
 		 \bottomrule 
    \end{tabular} 
    \caption{Standard deviation of risk group factors across bootstrap samples by risk group and method.} 
    \label{tab:c5_sd_rg_bootstrap} 
\end{table}

The CCF volatility as measured by the standard deviation is the lowest in the majority of the economic sectors. In the bootstrap samples, this applies to all but a few economic sectors, showing that the stabilizing effects of the CCF are more present as the data gets more volatile. In the tariff year data, the k-means methods are occasionally more stable for some economic sectors. These economic sectors likely do not experience a risk group change between the tariff years in the k-means methods. To assess which methods constructs more stable risk group allocations in general, we examine the transition matrices derived from all three approaches.

\noindent \\
\textbf{Risk group stability}
\noindent \\
To measure the stability of the risk group allocation across tariff years, we examine transition 
matrices. Similarly to transition matrices in Markov chains, a cell in the transition matrices show the 
proportion of economic sectors that start in risk group $a$ in tariff year $t$ and end up in risk group 
$b$ in tariff year $t+1$. Given that the risk groups form a partition, each row sums up to one. The 
higher the observed frequency in the diagonal of the transition matrix, the greater the stability of 
the clustering allocation across tariff years.
\par 
The transition matrix below shows that the CCF yields comparatively stable results for the odd-numbered 
risk groups (cf. Table~\ref{tab:tm_ccf}). For the middle risk group, the application of the 
multidimensional Bühlmann-Straub model ensures that the risks with a small statistical reliability are consistently 
close to global experience and are hence unlikely to switch risk groups. Conversely, risks with large 
volumes and claims experience that deviates from the global average tend to be consistently better or 
worse than the collective, thereby contributing to the observed stability in the extreme risk groups. 
\par
Similarly to the extreme risk groups, the even-numbered risk groups  contain risks with sufficiently 
large volumes to deviate from the global experience. In contrast to the extreme risk groups, however, 
these even-numbered risk groups share boundaries with two adjacent risk groups rather than one. As a 
result, they include twice as many risks for which only a small change in the claims experience can 
lead to a transition to a different risk group.
\begin{table}[H]
	\centering
	\begin{tabular*}{\textwidth}{@{\extracolsep{\fill}}lrrrrr}
		\toprule
		& \multicolumn{5}{c}{\textit{Credibility Clustering}} \\
		& \multicolumn{5}{c}{\textbf{Risk Group (Next Period)}} \\
		\cmidrule(lr){2-6}
		\textbf{Risk Group (Current)} & \textbf{1} & \textbf{2} & \textbf{3} & \textbf{4} & \textbf{5} \\
		\midrule
		1 & 73.6\% & 26.4\% & 0\% & 0\% & 0\% \\
		2 & 15.8\% & 50.6\% & 32.7\% & 0.9\% & 0\% \\
		3 & 0\% & 14.0\% & 66.9\% & 19.1\% & 0\% \\
		4 & 0\% & 0\% & 22.2\% & 60.7\% & 17.0\% \\
		5 & 0\% & 0\% & 0\% & 24.1\% & 75.9\% \\
		\bottomrule
	\end{tabular*}
		\caption{Transition matrix for the credibility clustering framework.}
	\label{tab:tm_ccf}
\end{table}
\par 
In contrast to the CCF, the most stable risk groups under the k-means methods are the first two groups 
(cf. Tables~\ref{tab:tm_kmwo} and~\ref{tab:tm_kmo}). These risk groups predominantly consist of risks 
with very few to no observed claims, resulting in a very favorable claims ratio. The remaining risk 
groups exhibit comparatively low stability.
\begin{table}[H]
	\centering
	\begin{tabular*}{\textwidth}{@{\extracolsep{\fill}}lrrrrr}
		\toprule
		& \multicolumn{5}{c}{\textit{K-Means (without Outliers)}} \\
		& \multicolumn{5}{c}{\textbf{Risk Group (Next Period)}} \\
		\cmidrule(lr){2-6}
		\textbf{Risk Group (Current)} & \textbf{1} & \textbf{2} & \textbf{3} & \textbf{4} & \textbf{5} \\
		\midrule
		1 & 84.7\% & 10.9\% & 4.3\% & 0\% & 0\% \\
		2 & 5.2\% & 67.8\% & 19.4\% & 6.6\% & 1\% \\
		3 & 0\% & 19.8\% & 57.8\% & 17.9\% & 4.5\% \\
		4 & 0\% & 0.8\% & 45.2\% & 29.1\% & 24.9\% \\
		5 & 0\% & 0\% & 3.3\% & 30.5\% & 66.2\% \\
		\bottomrule
	\end{tabular*}
		\caption{Transition matrix for the k-means method without outliers.}
	\label{tab:tm_kmwo}
\end{table}
\begin{table}[H]
	\centering
	\begin{tabular*}{\textwidth}{@{\extracolsep{\fill}}lrrrrr}
		\toprule
		& \multicolumn{5}{c}{\textit{K-Means (with Outliers)}} \\
		& \multicolumn{5}{c}{\textbf{Risk Group (Next Period)}} \\
		\cmidrule(lr){2-6}
		\textbf{Risk Group (Current)} & \textbf{1} & \textbf{2} & \textbf{3} & \textbf{4} & \textbf{5} \\
		\midrule
		1 & 83.5\% & 13\% & 1.7\% & 0\% & 1.8\% \\
		2 & 4.9\% & 80.6\% & 12\% & 2.4\% & 0\% \\
		3 & 0\% & 51.7\% & 34.7\% & 13\% & 0.7\% \\
		4 & 0\% & 0\% & 41\% & 48.8\% & 10.1\% \\
		5 & 0\% & 0\% & 15\% & 29.3\% & 55.7\% \\
		\bottomrule
	\end{tabular*}
		\caption{Transition matrix for the k-means method with outliers.}
	\label{tab:tm_kmo}
\end{table}
\par 
Overall, the CCF shows the most favorable temporal stability among the three explored methods. However, 
the general instability of the allocation -- evidenced by approximately a quarter of the economic 
sectors experiencing a tariff factor change greater than 10\% (cf. Table \ref{tab:c5_method_comparison}) -- is problematic in practice. This instability is largely driven by the risk group allocation 
instability of risks at the boundary between two risk groups, which is a result of using hard 
clustering techniques, such as k-means. Hence, as a potential improvement, the application of soft 
clustering techniques (such as fuzzy c-means~\cite{bezdek2013pattern} or Gaussian mixture 
models~\cite{YANG20123950}) could be explored. These methods may reduce instability caused by abrupt 
jumps, as they do not rely on hard cluster assignments. However, the present credibility-clustering framework does not directly accommodate soft memberships and would require an extension of the aggregation and credibility steps. 

\section{Conclusion}
\label{sec:conclusion}
The creation of risk groups is a vital part of the pricing process in the insurance industry. In group life insurance, this process is often done manually, which can lead to arbitrary decisions based on expert judgment. Unsupervised learning methods can be used to automate this process. However, the direct application of these methods (e.g. k-means) leads to non-viable results that neglect essential characteristics, such as the statistical reliability of the observations and the temporal stability of the solution. This paper has introduced the credibility clustering framework (CCF) to address these weaknesses.
\par
The CCF improves the viability of the results compared to the direct application of k-means (with and without outliers), showing: 
\begin{enumerate}
	\item Better performance regarding the consideration of the statistical reliability of the observations (see Section~\ref{sec:res_ss}).
	\item Better performance in terms of the stability of the risk group factors across different bootstrap samples (see Section~\ref{sec:res_v}).
	\item Better performance in the temporal stability (cf. Section~\ref{sec:res_ts}). 
\end{enumerate}
\par 
Notwithstanding these improvements, the credibility clustering framework still has some limitations, particularly in terms of temporal stability and the resulting abrupt tariff factor changes. As a direction for future research, the application of soft clustering methods could be explored to address the instability in the transition matrices. However, this introduces the additional challenge of adapting credibility models to soft risk group assignments, which requires further methodological development.

\section*{Acknowledgments}
The authors thank Milo\v{s} Radonji\'c and Martino Stoffel for their insightful discussions and feedback throughout the development of this paper.

\section*{Conflict of interest}
The authors declare that they have no potential conflicts of interest. The analysis was conducted with strict adherence to scientific objectivity and impartiality. At no point during the study did any conflicts of interest arise.
All data were processed in anonymized form to ensure that no individuals or organizations could be identified. Data selection and evaluation were based solely on scientific relevance and the objectives of the study. These findings reflect the authors’ independent scientific reasoning and personal assessments, and should not be construed as reflecting the interests or positions of the authors’ current employer or affiliated university.

\newpage
\bibliographystyle{unsrt}
\bibliography{references} 

\newpage
\appendix

\section{Additional Results}
\subsection{Sensitivity Analysis: Shrinkage Factor $\rho$}
\label{sec:app_rho}
Tables~\ref{tab:rho_avg_abs_change_rho} and~\ref{tab:rho_prop_change_rho} 
show how changes in the shrinkage factor $\rho$ affect the stability of 
tariff factors across tariff year changes. Increasing $\rho$ initially 
improves stability, particularly with respect to the proportion of changes 
exceeding 10\%. However, these improvements become less pronounced and 
largely plateau once $\rho$ exceeds 0.35. Overall, the impact of the 
shrinkage factor on the results is modest and does not alter the conclusions 
drawn in the main analysis.
\begin{table}[H] 
	\centering 
	\caption{Average absolute change by correlation parameter $\rho$.} 
	\label{tab:rho_avg_abs_change_rho} 
	\begin{tabular}{lcccccc} 
		\toprule \textbf{Method} & $\rho=0$ & $\rho=0.2$ & $\rho=0.35$ & $\rho=0.5$ & $\rho=0.75$ & $\rho=0.99$ \\ 
		\midrule CCF & 7.61\% & 7.21\% & 7.20\% & 7.02\% & 7.07\% & 7.02\% \\ 
		\bottomrule 
	\end{tabular} 
\end{table}

\begin{table}[H] 
	\centering 
	\caption{Proportion of changes exceeding 10\% by correlation parameter $\rho$.} 
	\label{tab:rho_prop_change_rho} 
	\begin{tabular}{lcccccc} 
		\toprule \textbf{Method} & $\rho=0$ & $\rho=0.2$ & $\rho=0.35$ & $\rho=0.5$ & $\rho=0.75$ & $\rho=0.99$ \\ 
		\midrule CCF & 35.95\% & 30.00\% & 23.57\% & 26.19\% & 23.10\% & 25.00\% \\ 
		\bottomrule 
	\end{tabular} 
\end{table}
\subsection{Sensitivity Analysis: Number of Cluster Centers $K$}
\label{sec:app_clus}
This section examines the sensitivity of the results to the choice of the number of cluster centers $K$ by comparing the CCF with the benchmark methods. Overall, the results for $K=3$ and $K=7$ confirm that the conclusions of the main analysis remain largely robust to the choice of the number of cluster centers, although the relative advantage of CCF becomes less uniform for K=7, particularly for tariff-year stability at the sector level.
\subsubsection{$K=3$}
\begin{figure}[htbp]
	\centering
	\includegraphics[width=\textwidth]{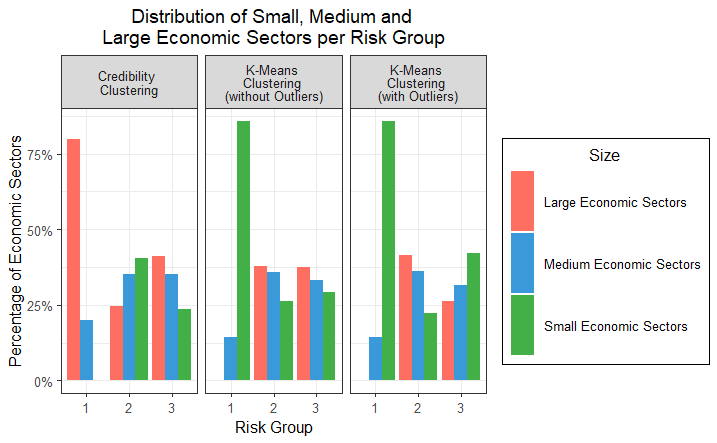}
	\caption{Figure showing the distribution of small (green), medium (blue) and large (red) economic sectors within the risk groups depending on the type of clustering used. Small, medium and large economic sectors correspond to the bottom, middle and top 33.3\% of economic sectors, as measured by their summed normalized volumes.}
	\label{fig:c3_es}
\end{figure}

Figure~\ref{fig:c3_es} shows that the CCF accounts for statistical reliability more effectively than the benchmark methods.
\par
\begin{figure}[htbp]
	\centering
	\includegraphics[width=\textwidth]{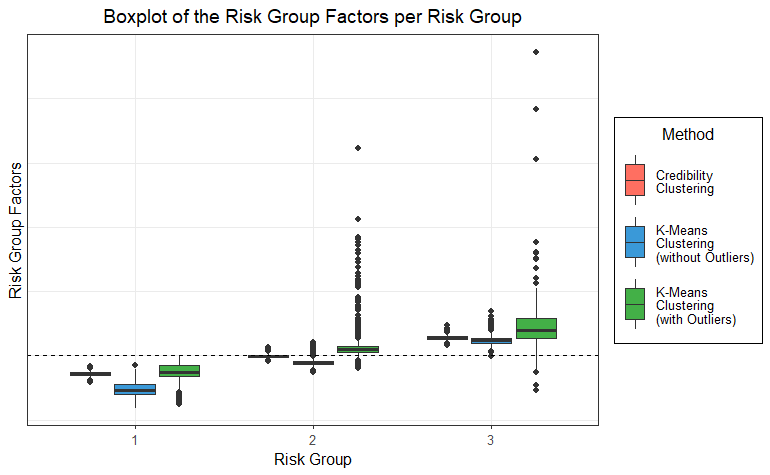}
	\caption{Boxplots of the risk group factors for risk groups one to three, comparing the CCF (red, left) with k-means clustering, both excluding (blue, middle) and including (green, right) outliers. The boxes represent the interquartile range (IQR), with whiskers extending up to 1.5 times the IQR. Dots indicate outliers beyond this range. The dashed horizontal line indicates the global risk experience (value of 1). The y-axis is omitted for confidentiality reasons.}
	\label{fig:c3_bs}
\end{figure}

Furthermore, Figure~\ref{fig:c3_bs} shows that all methods satisfy the viability criteria, with the CCF exhibiting the strongest overall performance, primarily due to its central alignment and stable risk group factors. However, the lower number of risk groups limits the achievable differentiation between risk group factors.
\par
Table~\ref{tab:c3_method_comparison} indicates that overall stability increases compared with the $K=5$ case. The CCF continues to yield the most stable results among the considered methods. This observation is corroborated by Tables~\ref{tab:c3_sd_rg_tariff_year} and~\ref{tab:c3_sd_rg_bootstrap}, where the standard deviations of the risk group factors are consistently lower for the CCF than for the benchmark approaches.
\begin{table}[H]
	\centering
	\small
	\begin{tabular}{lcc}
		\hline
		\multicolumn{1}{l}{\textbf{Method}} &
		\multicolumn{1}{c}{\makecell{\textbf{Average Absolute} \\ \textbf{Change}}} &
		\multicolumn{1}{c}{\makecell{\textbf{Proportion of Changes} \\ \textbf{$\geq$ 10\%}}} \\
		\hline
		CCF & 4.99\% & 11.90\% \\
		K-Means (without Outliers)& 10.71\% & 27.38\% \\
		K-Means (with Outliers) & 19.73\% & 56.43\% \\
		\hline
	\end{tabular}
	\caption{Average absolute change and proportion of tariff-factor changes exceeding 10\% across all tariff-year changes for the three methods.}
	\label{tab:c3_method_comparison}
\end{table}
\begin{table}[htbp] 
	\centering 
	\begin{tabular}{lccc} 
		\toprule & \multicolumn{3}{c}{\textbf{Risk Group}} \\ 
		\cmidrule(lr){2-4} \textbf{Method} & \textbf{1} & \textbf{2} & \textbf{3} \\ 
		\midrule Credibility Clustering & 0.024 & 0.015 & 0.020 \\ 
		K-Means (without Outliers) & 0.032 & 0.019 & 0.053 \\ 
		K-Means (with Outliers) & 0.124 & 0.066 & 0.160 \\ 
		\bottomrule 
	\end{tabular} 
	\caption{Standard deviation of risk group factors across tariff years by risk group and method.} 
	\label{tab:c3_sd_rg_tariff_year} 
\end{table} 

\begin{table}[htbp] 
	\centering 
	\begin{tabular}{lccc} 
		\toprule & \multicolumn{3}{c}{\textbf{Risk Group}} \\ 
		\cmidrule(lr){2-4} \textbf{Method} & \textbf{1} & \textbf{2} & \textbf{3} \\ 
		\midrule Credibility Clustering & 0.035 & 0.027 & 0.033 \\ 
		K-Means (without Outliers) & 0.105 & 0.066 & 0.068 \\ 
		K-Means (with Outliers) & 0.126 & 0.109 & 0.242 \\ 
		\bottomrule 
	\end{tabular} 
	\caption{Standard deviation of risk group factors across bootstrap samples by risk group and method.} 
	\label{tab:c3_sd_rg_bootstrap} 
\end{table}
\par
\begin{figure}[H]
	\centering
	\includegraphics[width=\textwidth]{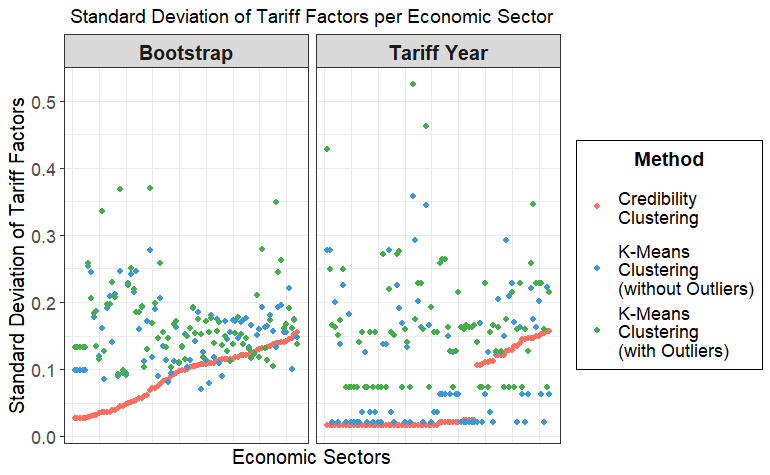}
	\caption{Graph comparing the standard deviation of the tariff factors across bootstrap samples and tariff year changes per economic sector and method. The data are ordered by the standard deviation of the CCF risk group factors. The x-axis labels are omitted due to confidentiality.}
	\label{fig:c3_sd}
\end{figure}

Lastly, the CCF yields the most stable tariff factors for the majority of economic sectors in both the bootstrap and tariff year analyses (cf. Figure~\ref{fig:c3_sd}).

\subsubsection{K=7}
\begin{figure}[H]
	\centering
	\includegraphics[width=\textwidth]{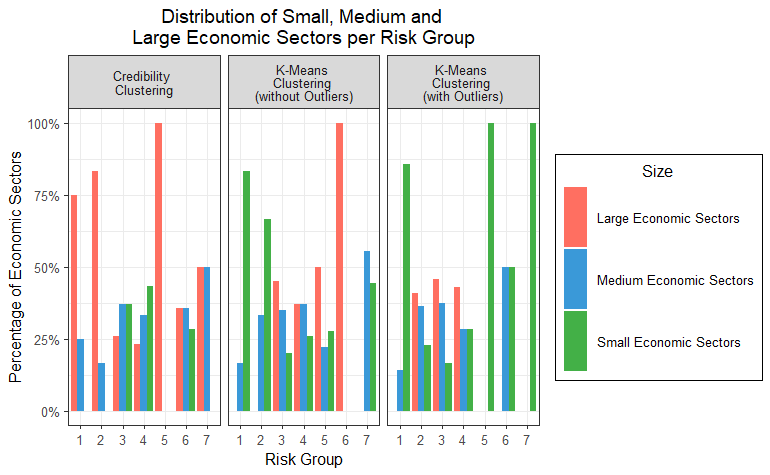}
	\caption{Figure showing the distribution of small (green), medium (blue) and large (red) economic sectors within the risk groups depending on the type of clustering used. Small, medium and large economic sectors correspond to the bottom, middle and top 33.3\% of economic sectors, as measured by their summed normalized volumes.}
	\label{fig:c7_es}
\end{figure}

Figure~\ref{fig:c7_es} shows that the CCF continues to account for statistical reliability more effectively than the benchmark methods, yielding a more credible allocation of economic sectors to the extreme risk groups.
\par
\begin{figure}[H]
	\centering
	\includegraphics[width=\textwidth]{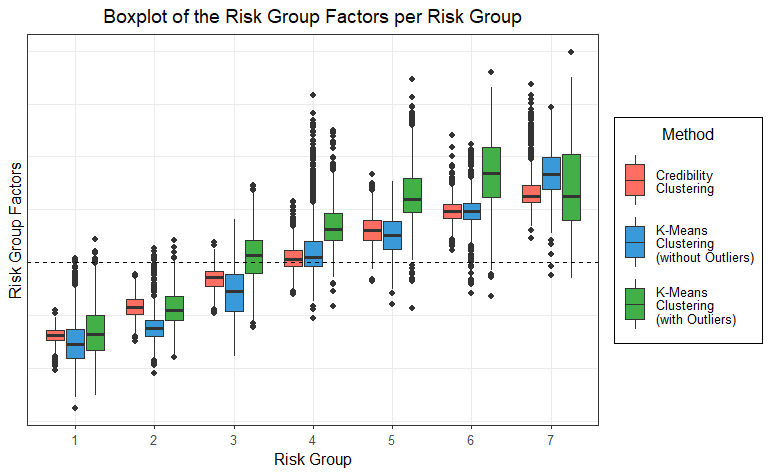}
	\caption{Boxplots of the risk group factors for risk groups one to seven, comparing the CCF (red) with k-means clustering, both excluding (blue) and including (green) outliers. The boxes represent the interquartile range (IQR), with whiskers extending up to 1.5 times the IQR. Dots indicate outliers beyond this range. The dashed horizontal line indicates the global risk experience (value of 1). The y-axis is omitted for confidentiality reasons.}
	\label{fig:c7_bs}
\end{figure}

Figure~\ref{fig:c7_bs} shows that the CCF satisfies the viability criteria, whereas the benchmark methods fail due to substantial overlap between the risk group factors. Although the differentiation between risk group factors is less pronounced for $K=7$ than for $K=5$, the resulting risk group structure remains sufficiently differentiated.
\par
The stability of the tariff factor changes is highest for the CCF, which continues to outperform the benchmark methods (cf. Table~\ref{tab:c7_method_comparison}). This result is further reflected in the stability of the risk group factors, where the CCF exhibits the lowest standard deviations for most risk groups in the tariff year data and for all risk groups in the bootstrap analysis (cf. Tables~\ref{tab:c7_sd_rg_tariff_year} and~\ref{tab:c7_sd_rg_bootstrap}).
\begin{table}[H]
	\centering
	\small
	\begin{tabular}{lcc}
		\hline
		\multicolumn{1}{l}{\textbf{Method}} &
		\multicolumn{1}{c}{\makecell{\textbf{Average Absolute} \\ \textbf{Change}}} &
		\multicolumn{1}{c}{\makecell{\textbf{Proportion of Changes} \\ \textbf{$\geq$ 10\%}}} \\
		\hline
		CCF & 6.35\% & 26.90\% \\
		K-Means (without Outliers)& 12.24\% & 40.48\% \\
		K-Means (with Outliers) & 11.43\% & 35.48\% \\
		\hline
	\end{tabular}
	\caption{Average absolute change and proportion of tariff-factor changes exceeding 10\% across all tariff-year changes for the three methods.}
	\label{tab:c7_method_comparison}
\end{table}
 \begin{table}[htbp] 
 	\centering 
 	\begin{tabular}{lccccccc} 
 		\toprule & \multicolumn{7}{c}{\textbf{Risk Group}} \\ 
 		\cmidrule(lr){2-8} \textbf{Method} & \textbf{1} & \textbf{2} & \textbf{3} & \textbf{4} & \textbf{5} & \textbf{6} & \textbf{7} \\ 
 		\midrule Credibility Clustering & 0.015 & 0.052 & 0.063 & 0.041 & 0.038 & 0.036 & 0.053 \\ 
 		K-Means (without Outliers) & 0.042 & 0.067 & 0.065 & 0.027 & 0.083 & 0.039 & 0.065 \\ 
 		K-Means (with Outliers) & 0.083 & 0.031 & 0.036 & 0.105 & 0.151 & 0.082 & 0.145 \\ 
 		\bottomrule 
 	\end{tabular} 
 	\caption{Standard deviation of risk group factors across tariff years by risk group and method.} 
 	\label{tab:c7_sd_rg_tariff_year} 
 \end{table} 
 
 \begin{table}[htbp] 
 	\centering 
 	\begin{tabular}{lccccccc} 
 		\toprule & \multicolumn{7}{c}{\textbf{Risk Group}} \\ 
 		\cmidrule(lr){2-8} \textbf{Method} & \textbf{1} & \textbf{2} & \textbf{3} & \textbf{4} & \textbf{5} & \textbf{6} & \textbf{7} \\ 
 		\midrule Credibility Clustering & 0.036 & 0.053 & 0.051 & 0.060 & 0.070 & 0.055 & 0.077 \\ 
 		K-Means (without Outliers) & 0.112 & 0.064 & 0.105 & 0.123 & 0.088 & 0.076 & 0.106 \\ 
 		K-Means (with Outliers) & 0.117 & 0.074 & 0.112 & 0.097 & 0.129 & 0.168 & 0.195 \\ 
 		\bottomrule 
 	\end{tabular} 
 	\caption{Standard deviation of risk group factors across bootstrap samples by risk group and method.} 
 	\label{tab:c7_sd_rg_bootstrap} 
\end{table}
\par
\begin{figure}[H]
	\centering
	\includegraphics[width=\textwidth]{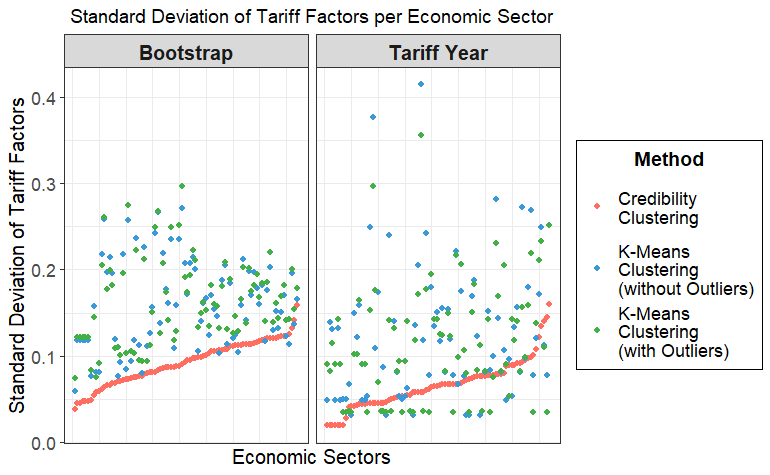}
	\caption{Graph comparing the standard deviation of the tariff factors across bootstrap samples and tariff year changes per economic sector and method. The data are ordered by the standard deviation of the CCF risk group factors. The x-axis labels are omitted due to confidentiality.}
	\label{fig:c7_sd}
\end{figure}

Lastly, the stability of the tariff factors at the economic-sector level, as measured by the standard deviation, is highest for the CCF in most sectors for the bootstrap data (cf. Figure~\ref{fig:c7_sd}). In the tariff year data, however, the CCF no longer consistently outperforms the benchmark methods. A possible explanation is that increasing the number of risk groups raises the likelihood of transitions between adjacent risk groups, resulting in higher tariff-factor volatility.

\end{document}